\documentclass[conference]{IEEEtran}
\IEEEoverridecommandlockouts
\usepackage{cite}
\usepackage{amsmath,amssymb,amsfonts}
\usepackage{algorithmic}
\usepackage{graphicx}
\usepackage{textcomp}
\usepackage{xcolor}

\usepackage{booktabs}
\usepackage[linesnumbered,ruled,vlined]{algorithm2e}
\usepackage{multirow}
\usepackage{enumitem}
\usepackage{mathrsfs}
\usepackage{newtxtext, newtxmath}
\newcommand{\stitle}[1]{\noindent \textbf{#1}}
\usepackage{etoolbox}%
\providetoggle{beginFlag}
\settoggle{beginFlag}{true}

\makeatletter
\patchcmd{\@algocf@start}{-1.5em}{-0.67em}{}{} %
\makeatother
\makeatletter
\let\NAT@parse\undefined
\makeatother
\usepackage{hyperref}  %

\def\BibTeX{{\rm B\kern-.05em{\sc i\kern-.025em b}\kern-.08em
    T\kern-.1667em\lower.7ex\hbox{E}\kern-.125emX}}
\begin{document}
\bstctlcite{IEEEexample:BSTcontrol}

\title{BQSched: A Non-intrusive Scheduler for Batch Concurrent Queries via Reinforcement Learning
\thanks{\IEEEauthorrefmark{1}Corresponding authors: Jiannan Wang (jnwang@sfu.ca) and Jun Gao (gaojun@pku.edu.cn)}}

\author{
\IEEEauthorblockN{Chenhao Xu\IEEEauthorrefmark{2}, Chunyu Chen\IEEEauthorrefmark{3}, Jinglin Peng\IEEEauthorrefmark{4}, Jiannan Wang\IEEEauthorrefmark{4}\IEEEauthorrefmark{3}\IEEEauthorrefmark{1}, and Jun Gao\IEEEauthorrefmark{2}\IEEEauthorrefmark{1}}
\IEEEauthorblockA{
\textit{\IEEEauthorrefmark{2}Key Laboratory of High Confidence Software Technologies, CS, Peking University, Beijing, China,}\\
\textit{\IEEEauthorrefmark{3}Simon Fraser University, Burnaby, Canada, \IEEEauthorrefmark{4}Huawei Cloud, Beijing, China}
}
}

\maketitle

\begin{abstract}
Most large enterprises build predefined data pipelines and execute them periodically to process operational data using SQL queries for various tasks. A key issue in minimizing the overall makespan of these pipelines is the efficient scheduling of concurrent queries within the pipelines. Existing tools mainly rely on simple heuristic rules due to the difficulty of expressing the complex features and mutual influences of queries. The latest reinforcement learning (RL) based methods have the potential to capture these patterns from feedback, but it is non-trivial to apply them directly due to the large scheduling space, high sampling cost, and poor sample utilization.

Motivated by these challenges, we propose \textbf{BQSched}, a non-intrusive Scheduler for Batch concurrent Queries via reinforcement learning. Specifically, BQSched designs an attention-based state representation to capture the complex query patterns, and proposes IQ-PPO, an auxiliary task-enhanced proximal policy optimization (PPO) algorithm, to fully exploit the rich signals of Individual Query completion in logs. Based on the RL framework above, BQSched further introduces three optimization strategies, including adaptive masking to prune the action space, scheduling gain-based query clustering to deal with large query sets, and an incremental simulator to reduce sampling cost. To our knowledge, BQSched is the first non-intrusive batch query scheduler via RL. Extensive experiments show that BQSched can significantly improve the efficiency and stability of batch query scheduling, while also achieving remarkable scalability and adaptability in both data and queries. For example, across all DBMSs and scales tested, BQSched reduces the overall makespan of batch queries on TPC-DS benchmark by an average of 34\% and 13\%, compared with the commonly used heuristic strategy and the adapted RL-based scheduler, respectively. The source code of BQSched is available at \url{https://github.com/chxu2000/BQSched}.
\end{abstract}


\section{Introduction}
\label{sec:intro}
Most large enterprises predefine complex pipelines and execute them periodically to extract and transform operational data from various systems, including centralized, distributed, and cloud DBMSs, for tasks like decision-making and transaction monitoring \cite{therrien2020critical, goodhope2012building, o2015industrial, de2015crispred, klievink2012enhancing}. These pipelines consist of jobs with specific execution orders and complex internal queries \cite{raj2020modelling}, and thus consume substantial computational resources when executed periodically. For example, 80\% of queries in Amazon Redshift's workloads are found to be repeated \cite{nathan2024intelligent}. It is highly beneficial to find a strategy that can minimize the overall makespan, i.e., the total execution time, of these pipelines without additional hardware or service expenditure.

It is feasible to speed up pipeline execution through an efficient scheduling strategy. First, different queries may have different preferences for computational resources. For example, some of the queries in TPC-DS \cite{nambiar2006making} benchmark are I/O-intensive, while others mainly require CPU resources \cite{poess2007you}. If we run the I/O-intensive and CPU-intensive queries concurrently, it can avoid possible resource contention among concurrent queries. Second, all queries share the same data buffer in one DBMS, indicating that one query may reuse the data loaded by others \cite{effelsberg1984principles}. We can then improve the overall efficiency by concurrently executing queries that share resources. Third, slow queries, especially when scheduled late, may significantly extend the overall makespan. Such a long-tail query problem can be intuitively mitigated by scheduling the slow queries early.

Despite its feasibility and benefits, recent research on pipeline scheduling is limited, while some works focus on scheduling distributed jobs for data processing \cite{mao2019learning} and deep learning (DL) \cite{peng2021dl2, chen2022rifling, xing2023dual}. This is partly due to the complex features and mutual influences of queries. Specifically, SQL queries can be quite complex, with some TPC-DS queries spanning hundreds of lines. In addition, there is not only resource contention, but also resource sharing, such as shared data loading among queries, which is not common among distributed jobs.

This paper focuses on a key subproblem in pipeline execution, i.e., \textit{the scheduling of batch concurrent queries} (referred to as \textit{batch queries} hereafter) that can be invoked concurrently without dependency relationships. These queries can be extracted from a single job without inter-query dependencies, or from multiple jobs without inter-job dependencies. Obviously, efficiently scheduling these queries can accelerate the entire pipeline execution. Therefore, we abstract the core problem in pipeline execution as follows: Given an $n$-sized batch query set $S$ whose queries can be executed concurrently on connections $C$ to an underlying DBMS, how to determine the execution order, database connection, and running parameters for each query to reduce the overall makespan of $S$. Here, we set $\left|C\right|$ empirically to fully utilize the resources of the DBMS.

Capturing the complex features and mutual influences of queries is the basis for batch query scheduling. Most existing tools ignore these complexities and directly take simple heuristics based on runtime statistics. For example, DBT \cite{cyr2023introduction}, a popular data pipeline building tool, follows a simple first in, first out (FIFO) strategy, whose performance depends entirely on the order of input queries. Maximum cost first (MCF) is another strategy that first extracts the execution costs of individual queries from logs, and then schedules the slowest query first to alleviate the long-tail query problem. However, these heuristic strategies do not consider the resource sharing and contention among queries, leading to significant optimization potential in their scheduling plans. With the rapid advancement of AI4DB techniques, recent query representation methods \cite{zhao2022queryformer, sun2019end, marcus2021bao} convert physical plans into embeddings to capture key features of queries based on feedback from downstream tasks. Such methods provide a promising approach to modeling the complex features and mutual influences of queries.

Batch query scheduling can be formulated as a sequential decision problem, making it suitable for solving with RL \cite{mnih2013playing, silver2016mastering, silver2017mastering}. RL aims to optimize sequential decisions by interacting with the environment, where an agent selects an action at each step based on the current state to maximize the cumulative reward \cite{arulkumaran2017deep}. In our scenario, the learned representation of batch queries naturally serves as the state in RL, and the action of the scheduling agent is to select a query along with its running parameters and submit it to a DBMS connection, while the reward is designed based on the objective of minimizing the overall makespan (e.g., using the negative value of makespan as the reward). In fact, RL schedulers have been investigated in different applications, such as job scheduling for data processing \cite{mao2019learning} and DL \cite{peng2021dl2, chen2022rifling, xing2023dual} clusters, and operator-level query scheduling in database systems \cite{sabek2022lsched}. Inspired by these successes, we employ RL in our scenario to implicitly capture the complex query patterns based on feedback from the underlying DBMS, and thus learn an efficient scheduling strategy for batch queries.

From another perspective of how a scheduler interacts with the underlying system, most existing works follow an intrusive approach. For example, LSched \cite{sabek2022lsched} schedules plan operators based on white-box features, thus requiring integration with specific open-source DBMSs. Other works build a holistic plan for batch queries by extending join operators \cite{makreshanski2016mqjoin, makreshanski2018many} or physical plans \cite{giannikis2012shareddb, giannikis2014shared} to multiple queries, which can reduce redundant computation. However, these intrusive methods are tightly coupled with the query execution component, e.g., changing the behavior of plan operators. Therefore, they have to be integrated into the source code of the DBMS, which is generally unavailable, especially for commercial DBMSs. In contrast, non-intrusive schedulers treat the underlying systems as black boxes and adaptively learn scheduling strategies across centralized, distributed, and cloud DBMSs, despite their different internal \linebreak implementations, such as resource management policies.

This paper aims to employ RL to learn efficient scheduling strategies that minimize the overall makespan of batch queries in a non-intrusive way. However, before learning an efficient strategy, we need to address the following challenges, most of which \linebreak are inherent to RL but even more pronounced in our context.

\stitle{Complex Query Patterns.} It is a prerequisite for RL to capture the complex patterns of batch queries, including not only complicated features and diverse resource requirements at different execution stages, but also complex mutual influences, such as resource sharing and contention, during concurrent execution. We cannot simply represent such interactions using static features and information extracted from logs due to the flexible combination patterns of concurrent queries. Therefore, the learned query representation methods should be incorporated and extended in our scheduler.

\stitle{Large Scheduling Space.} As a common issue in RL \cite{dulac2021challenges}, the performance degradation from a large search space is even more serious in our context, where the space of scheduling plans grows exponentially with the number of batch queries and running parameters. In our scenario, the number of batch queries can range from 100 to 1,000, as these queries are extracted from predefined pipelines rather than arriving randomly \cite{mao2019learning, sabek2022lsched}. Although LSched \cite{sabek2022lsched} tests the performance under batch queries, the number of concurrent queries does not exceed 100. Thus, our scheduler requires special designs for the scheduling space to avoid high training costs and poor policy stability.

\stitle{High Sampling Cost and Poor Sample Utilization.} RL often requires plenty of samples to converge \cite{dulac2021challenges}, while sampling an episode, i.e., a complete sequence of batch query scheduling, is costly in our scenario. These factors lead to an urgent need to generate and utilize samples more efficiently. For sample generation, Decima \cite{mao2019learning} lowers sampling cost via a Spark \cite{zaharia2010spark} simulator with access to profiling information and the job run time characteristics. We follow a similar idea, but our simulator design is more difficult due to the complex query patterns and the requirement for non-intrusiveness. For sample utilization, existing RL schedulers \cite{mao2019learning, sabek2022lsched} aim to minimize average job completion time and can thus fully utilize the collected samples by exploiting local real-time feedback, such as individual query latencies. In contrast, our objective is to minimize the overall makespan of batch queries, leading to only one feedback per sequence for typical RL algorithms to exploit. Given that the rich signals of individual query completion are wasted, it is expected \linebreak to accelerate policy learning if these signals can be utilized.

\begin{figure*}[t]
    \centering
    \includegraphics[width=.908\linewidth]{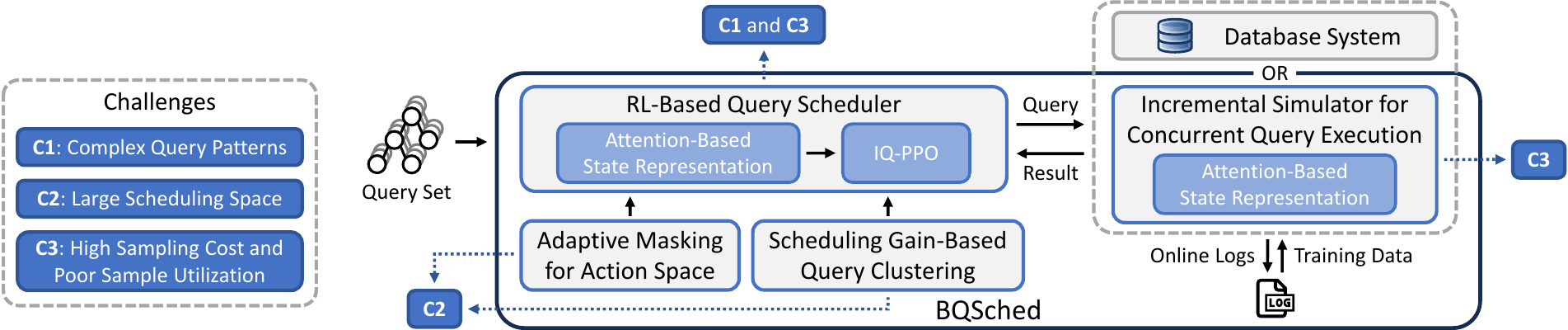}\vspace{-5.21pt}
    \caption{BQSched system overview.}
    \label{fig:system_overview}\vspace{-15.2pt}
\end{figure*}

In this paper, we propose \textbf{BQSched}, a non-intrusive \textbf{Sched}uler for \textbf{B}atch concurrent \textbf{Q}ueries via reinforcement learning, with the following contributions:
\begin{enumerate}[wide, labelwidth=!, labelindent=0pt]
    \item We devise an RL-based non-intrusive scheduler for batch queries. To capture the complex query patterns, we combine the execution plan encoded by QueryFormer \cite{zhao2022queryformer} with the running state features to represent each individual query, and model the complex mutual influences among queries through multi-head attention \cite{vaswani2017attention}. Moreover, we introduce IQ-PPO, an auxiliary task-enhanced PPO \cite{schulman2017proximal} algorithm that exploits the rich signals of Individual Query completion to improve sample utilization.
    \item We further introduce three optimization strategies: (i) We incorporate adaptive masking to prune the action space by masking inefficient parameter configurations, which reduces the scheduling space with limited impact on the overall execution efficiency. (ii) Observing that the concurrently executed queries in an efficient scheduling always illustrate some performance gain, we propose a scheduling gain-based query clustering method to deal with large query sets by scheduling at the cluster level. (iii) We design an incremental simulator to reduce sampling cost by mimicking feedback from the underlying DBMS. The \linebreak simulator shares the state representation of our decision model.
    \item We evaluate BQSched on various benchmarks, databases, data scales, and query scales. As there is no RL-based non-intrusive scheduler for batch queries, we adapt LSched \cite{sabek2022lsched}, the state-of-the-art RL-based query scheduler, to our context. The results show that BQSched not only significantly outperforms the commonly used heuristic strategies and the adapted RL-based scheduler in both efficiency and stability, but also achieves remarkable scalability and adaptability under various data and query scales. For example, across all DBMSs and scales tested, BQSched reduces the overall makespan of batch queries on TPC-DS \cite{nambiar2006making} benchmark by an average of 34\% and 13\%, compared with FIFO and LSched, respectively. Moreover, pre-training BQSched on our simulator reduces the overall training time by about 80\% on TPC-DS without sacrificing the performance.
\end{enumerate}

\section{Problem and Method Overview}
In this section, we present the formulation of batch query scheduling and an overview of our BQSched's framework.

\subsection{Problem Formulation}\label{subsec:problem_formulation}
\stitle{Problem Definition.} Given a set $S$ of $n$ batch queries, a DBMS with connections $C$ whose cardinality $\left|C\right|$ is set by default to maximize execution efficiency based on hardware resources, and query running parameters $R$ with each $r_j \in R$ having $k_j$ configurations to adjust the degree of parallelism, memory limit, etc., we determine the execution order, the database connection $c_i \in C$, and each running parameter $r_j^i \in R^i$ for each query $q_i \in S$ to form a scheduling plan. With the execution of $S$, we get the finish time $T_i$ of each $q_i$ based on feedback from the DBMS. After all $q_i \in S$ finish, the overall makespan $t^{ov}$ of $S$ can be expressed as the latest finish time $\max(T_i)$ for all $q_i \in S$.

Given that each scheduling by the same strategy in the same scenario may produce different $t^{ov}$ due to the complex interactions among concurrent queries, we define two metrics to evaluate the performance of a strategy, including the average makespan $\overline{t}^{ov}$ and the standard deviation $\sigma^{ov}$ over multiple rounds of scheduling, e.g., $m=5$, under the same settings:
\begingroup
\setlength{\abovedisplayskip}{5pt}
\setlength{\belowdisplayskip}{5pt}
\begin{align*}
    \overline{t}^{ov} = \frac{1}{m}\sum_{i=1}^m{t^{ov}_i}, \ \sigma^{ov} = \sqrt{\frac{1}{m}\sum_{i=1}^m{\left(t^{ov}_i-\overline{t}^{ov}\right)^2}}
\end{align*}
\endgroup

This paper aims to develop a scheduler that learns strategies with shorter $\overline{t}^{ov}$ and smaller $\sigma^{ov}$, thereby enhancing efficiency and stability. In addition, the scheduler needs to maintain good scalability as the data and query scales increase. The learned strategies must also be adaptable to relatively small variations in data and query sets.

\stitle{Complexity Analysis.} We analyze the problem's complexity by enumerating potential scheduling plans for the batch queries. There will be $O\left(n! \cdot \left|C\right|^n\right)$ different plans as we can place the batch queries in any order on any connection. Considering the $k_j$ configurations of each $r_j$ in total $\left|R\right|$ running parameters, the number of valid scheduling plans will further increase exponentially to $O\left(n! \cdot \left(\left|C\right| \cdot \prod_{j=1}^{\left|R\right|}{k_j}\right)^n\right)$. Among these, some plans significantly reduce the overall makespan by utilizing resources efficiently and avoiding the long-tail query problem.

This paper follows a reasonable heuristic rule to simplify the problem and reduce its complexity. During the scheduling process, we improve resource utilization by keeping all the $\left|C\right|$ connections busy. That is, we select and submit the next query to execute to connection $c_i$ once the previous query on $c_i$ finishes. The above simplification reduces the problem's complexity from $O\left(n! \cdot \left(\left|C\right| \cdot \prod_{j=1}^{\left|R\right|}{k_j}\right)^n\right)$ to $O\left(n! \cdot \left(\prod_{j=1}^{\left|R\right|}{k_j}\right)^n\right)$ by avoiding the selection of a specific connection during scheduling. However, the simplified problem is still too complex to be solved using off-the-shelf heuristic or RL-based scheduling methods.

\begin{figure*}[htbp]
    \centering
    \includegraphics[width=.961\linewidth]{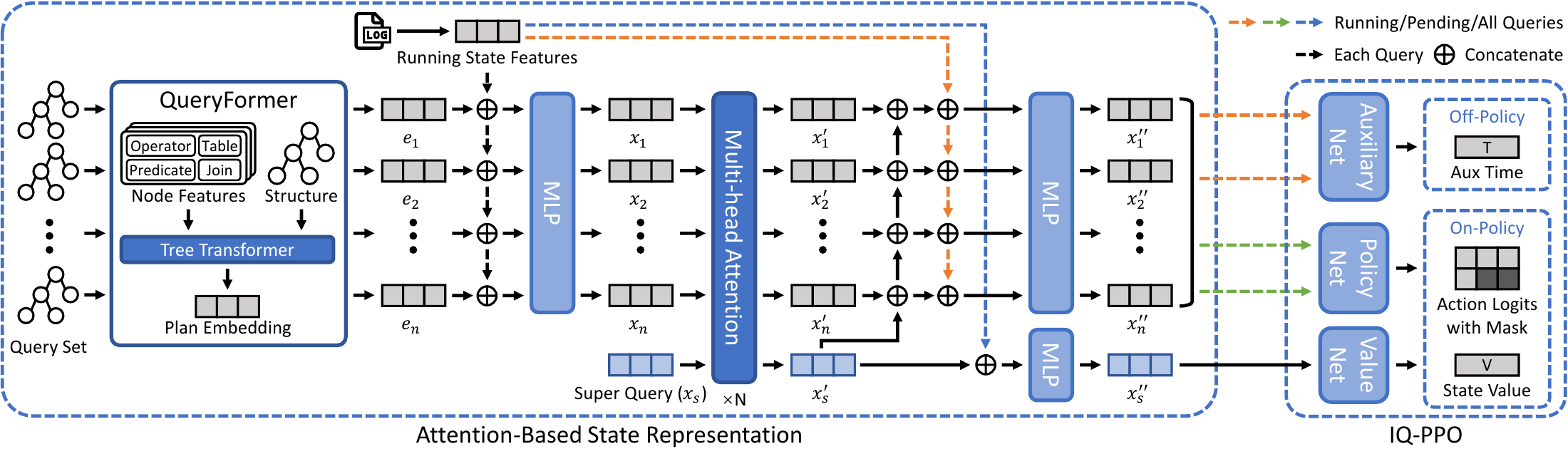}\vspace{-5pt}
    \caption{RL-based query scheduler.}
    \label{fig:rl_scheduler}\vspace{-14pt}
\end{figure*}

\subsection{System Overview}

Figure \ref{fig:system_overview} gives an overview of BQSched and its internal data flow. The system consists of four components: an RL-based query scheduler, adaptive masking for action space, scheduling gain-based query clustering, and an incremental simulator for concurrent query execution. The last three can be viewed as optimization strategies for the RL scheduler.

The RL-based query scheduler schedules batch queries based on their execution state and optimizes the strategy via RL. First, we combine the execution plan encoded by QueryFormer \cite{zhao2022queryformer} with the running state features to represent each query. Given the complex mutual influences among queries, we compute the final representation for each query and the overall state representation through multi-head attention. Based on these, the scheduler selects the next query to execute along with its running parameters. To optimize the policy and value networks, we propose IQ-PPO, which introduces an auxiliary time prediction task \linebreak to fully exploit the rich signals of individual query completion.

The adaptive masking masks inefficient parameter configurations for queries to prune the action space. Given that different queries have various preferences for computational resources, we incorporate the query performance under various configurations to reduce the search space. Specifically, based on the efficiency improvements brought by allocating more resources, we mask configurations, such as allocating more CPU resources \linebreak to I/O-intensive queries, to improve exploration efficiency.

The query clustering clusters batch queries based on their scheduling gains to deal with large query sets. As the scheduling space grows exponentially with the number of queries, we transform the scheduling granularity from queries to clusters to reduce the search space. First, we define and calculate the scheduling gain between two queries, i.e., the performance gain brought by executing them concurrently. Then, we perform agglomerative clustering on the queries and extend the RL scheduler to support cluster-level scheduling.

The incremental simulator simulates feedback from the underlying DBMS to reduce sampling cost. As it is costly for the DBMS to execute each concurrent query, we use the rich signals from logs to train the simulator to simulate feedback on query finish times by predicting the earliest query to finish and its finish time among all concurrent queries. Specifically, we extract query completion signals from offline logs produced by historical executions to train the simulator, and extract incremental data from online logs generated by more recent executions to fine-tune the simulator, adapting it to potential changes in the underlying data and queries. For policy learning, we pre-train the scheduler with samples collected from the simulator and fine-tune it using real-world data. Such a two-phase training paradigm significantly reduces the high sampling cost during training directly on the DBMS.

\section{RL-based Query Scheduler}
The primary goal of BQSched is to learn a scheduling strategy that can select the next query to execute along with its running parameters (action) based on the current state of batch queries (state) to reduce the overall makespan (reward). In this section, we introduce the major designs for batch query scheduling within our RL framework, including attention-based state representation and IQ-PPO, as shown in Figure \ref{fig:rl_scheduler}.

\subsection{Attention-Based State Representation}\label{subsec:attention_based_state_representation}
State representation is the basis for scheduling strategies. In our context, the state should capture the complex patterns of all batch queries, including their physical plans, running status, and mutual influences, to provide necessary information for decision-making. Therefore, we first extract useful features from each query. Then, we employ multi-head attention to capture the mutual influences among queries, thus obtaining the final representations for both individual queries and the global state. The attention mechanism also supports variable-length batch query inputs, which endows the learned policy with certain generalization abilities. We detail these two steps as follows.

\stitle{Single Query Representation.} For each query, its execution plan covers relevant tables, operators, etc., which reveals potential resource sharing and contention among queries. Moreover, some running state features, including running parameters and elapsed time, are not included in the plan but play an important role in identifying the state of queries. Therefore, the single query representation should capture both kinds of features.

We use QueryFormer \cite{zhao2022queryformer}, a state-of-the-art tree Transformer \cite{vaswani2017attention}, to encode tree-structured execution plans from individual nodes to the entire plan, as shown in Figure \ref{fig:rl_scheduler}. For node features, QueryFormer extracts operators, predicates, joins, and tables, which are crucial for identifying resource sharing and contention. For example, queries accessing the same tables may share data buffer, similar predicates may use the same indexes, while complex operators may lead to contention for CPU or I/O resources. QueryFormer also uses histograms and samples to bring database statistics into predicate encoding, generalizing the representation to potential changes in the underlying data distribution, which is also required in our context.

For the overall plan representation, QueryFormer introduces a super node directly connected to all other nodes, and encodes the entire plan with a tree Transformer, which models structural information via height encoding and tree-bias attention. Such information, including dependencies and information flow between operators, reveals resource usage at different stages of the query, which is also essential for modeling the mutual influences. Finally, QueryFormer takes the super node embedding as the plan embedding, thereby gathering features from all other nodes.

We further extract running state features $f_i$ for each query $q_i$, including the execution status $s_i$ (pending/running/finished), running parameters $R^i$, elapsed time $t_i$, and average execution time $\overline{t}_{i\vert R^i}$ under the given $R^i$, among which $\overline{t}_{i\vert R^i}$ is extracted from logs. These features are crucial for the scheduler to identify the resource allocation and execution stage of each query. We use one-hot encoding for discrete features such as $s_i$ and $R^i$, normalize continuous features such as $t_i$ and $\overline{t}_{i\vert R^i}$, and concatenate all features to get $f_i$:
\begingroup
\setlength{\abovedisplayskip}{2.7pt}
\setlength{\belowdisplayskip}{2.7pt}
\begin{align*}
    f_i = s_i \mathbin\Vert R^i \mathbin\Vert t_i \mathbin\Vert \overline{t}_{i\vert R^i}
\end{align*}
\endgroup
where $\mathbin\Vert$ represents the concatenation between features.

To obtain the representation $x_i$ for each query $q_i$, we concatenate the execution plan embedding $e_i$ encoded by QueryFormer with the running state features $f_i$, and then pass them through a multilayer perceptron (MLP) to further extract useful features:
\begingroup
\setlength{\abovedisplayskip}{5pt}
\setlength{\belowdisplayskip}{5pt}
\begin{align*}
    x_i = \left(\sigma \cdot \mathrm{Linear}\right)^\mathscr{m} \left(e_i \mathbin\Vert f_i\right)
\end{align*}
\endgroup
where $\sigma$ denotes a non-linear activation function (e.g., TanH), and $\mathscr{m}$ represents the number of the MLP layers.

\stitle{State Representation.} Based on the single query representations, which contain information revealing potential resource sharing and contention, we employ multi-head attention to model the complex mutual influences among queries. Similar to the [CLS] token in BERT \cite{kenton2019bert} and the super node in QueryFormer \cite{zhao2022queryformer}, we introduce an artificial super query with learnable initial features to gather key features from all batch queries and thus capture the global state for these queries. Then, we integrate the super query embedding encoded by the attention layers with the original query features to obtain the global state representation. In addition, we update the batch query representations similarly to support future decision-making.

As shown in Figure \ref{fig:rl_scheduler}, we feed the super query embedding $x_s$ and the batch query representations together into the attention layers for full interaction. The attention mechanism computes attention scores between queries in the batch query set, which allows each query to adaptively aggregate information based on the influence of other queries in the batch, thus considering the mutual influences among queries and enabling the super query to capture the global state. Moreover, multi-head attention enhances this by computing multiple sets of attention scores, which helps to capture various types of mutual influences, such as sharing and contending for CPU, I/O, and buffer resources.

Specifically, each attention layer contains two sub-layers: a multi-head attention (MHA) layer and a query-wise feed-forward (FF) layer. Moreover, we add a skip connection \cite{he2016deep} and batch normalization (BN) \cite{ioffe2015batch} for each sub-layer:
\begingroup
\setlength{\abovedisplayskip}{5pt}
\setlength{\belowdisplayskip}{5pt}
\begin{align*}
    \hat{x}_i &= \mathrm{BN}^\mathscr{l}\left(x_i^{\left(\mathscr{l}-1\right)} + \mathrm{MHA}_i^\mathscr{l}\left(x_1^{\left(\mathscr{l}-1\right)}, ..., x_n^{\left(\mathscr{l}-1\right)}, x_s^{\left(\mathscr{l}-1\right)}\right)\right) \\
    x_i^{\left(\mathscr{l}\right)} &= \mathrm{BN}^\mathscr{l}\left(\hat{x}_i + \mathrm{FF}^\mathscr{l}\left(\hat{x}_i\right)\right)
\end{align*}
\endgroup
where $\mathscr{l}$ is the layer index, $\hat{x}_i$ is the intermediate result, and $x_i^{\left(\mathscr{l}\right)}$ is $q_i$'s representation after $\mathscr{l}$ attention layers. Moreover, we update the super query embedding $x_s^{\left(\mathscr{l}\right)}$ in the same way.

Based on the super query embedding $x'_s$ and the representation $x'_i$ of each query $q_i$ encoded by the attention layers, we integrate the original query features to obtain the final representations for policy learning. Specifically, we concatenate $x'_s$ with the running state features of all queries $S$ by a skip connection, and compute the global state representation $x''_s$ with an MLP:
\begingroup
\setlength{\abovedisplayskip}{5pt}
\setlength{\belowdisplayskip}{5pt}
\begin{align*}
    x''_s = \left(\sigma \cdot \mathrm{Linear}\right)^\mathscr{n} \left(x'_s \mathbin\Vert \left(\mathbin\Vert_{j : q_j \in S}{\left(f_j\right)}\right) \right)
\end{align*}
\endgroup

To support decision-making, we update each query's representation similarly. We first concatenate $x'_i$ with $x'_s$ to incorporate global information. Then, considering the influence of concurrent queries $S_c$, we introduce their running state features and compute the final representation $x''_i$ with another MLP:
\begingroup
\setlength{\abovedisplayskip}{5pt}
\setlength{\belowdisplayskip}{5pt}
\begin{align*}
    x''_i = \left(\sigma \cdot \mathrm{Linear}\right)^\mathscr{o} \left(x'_i \mathbin\Vert x'_s \mathbin\Vert \left(\mathbin\Vert_{j : q_j \in S_c}{\left(f_j\right)}\right) \right)
\end{align*}
\endgroup

\subsection{Policy Learning with IQ-PPO}
In this subsection, we propose IQ-PPO, an auxiliary task-enhanced PPO algorithm that exploits the rich signals of Individual Query completion to improve sample utilization.

\stitle{Backbone and Issue.} Based on the state representation, we need to learn a policy to select the next query to execute and its running parameters. Among various RL algorithms, we choose PPO \cite{schulman2017proximal} as our backbone for its good performance and ease of use. PPO contains two components based on the shared state representation, where the actor learns a policy guided by the critic to maximize the cumulative reward. PPO can also reuse samples \linebreak generated by previous policies that are close to the current policy.

However, although PPO allows some sample reuse, its restriction on the difference between behavior and current policies still limits sample utilization, leading to the RL policy requiring plenty of samples to converge. This issue is particularly pronounced in batch query scheduling, where sampling a complete scheduling sequence is costly. Thus, it is crucial to fully utilize \linebreak the collected scheduling sequences to accelerate RL training.

Given that excessive sample reuse may lead to unstable policy updates, PPG \cite{cobbe2021phasic} attempts to improve sample utilization by introducing an auxiliary value prediction task to the policy network, as value function optimization often tolerates more sample reuse. However, the ground truth values used for auxiliary training, which are estimated via generalized advantage estimation \cite{schulman2015high}, may not accurately reveal the real state values. Other works \cite{jaderberg2017reinforcement, lee2019making} design domain-specific auxiliary tasks, which do not directly affect policy or value learning, but guide the learning \linebreak of shared state representations to improve sample utilization.

\stitle{Auxiliary Task in IQ-PPO.} Inspired by the ideas above, we incorporate another auxiliary task designed for batch query scheduling into the basic PPO algorithm.

Unlike typical RL algorithms that rely on overall makespans as sparse feedback, our auxiliary task aims to assist in state representation learning and improve sample utilization by exploiting the rich signals of individual query completion. During the policy learning process, the primary goal is to reduce the overall makespan of batch queries, which leads to the scheduler having only one sparse signal as feedback at the end of each time-consuming scheduling. In contrast, assuming that there are $n$ batch queries executed in one round, there will be $n$ individual query completion signals in logs, and we can extract $n$ sets of concurrent queries along with their execution states to predict their individual finish times. These rich signals are wasted in the basic RL training, but can be utilized by reusing the state representation in Section \ref{subsec:attention_based_state_representation}, as the representation of each query can be used to predict its finish time.

To fully exploit the individual query completion signals, we propose IQ-PPO, which incorporates an auxiliary task that predicts the finish time of each concurrent query based on the corresponding shared representation. By leveraging the underutilized real signals from the DBMS in logs, the auxiliary task can effectively assist in learning the state representations, thereby improving sample utilization.

\stitle{Overall Policy Learning.} As shown in Algorithm \ref{alg:iq-ppo}, IQ-PPO divides the training process into two alternating phases and performs auxiliary training every few iterations of the original PPO training to avoid interfering with PPO. During the PPO phase (lines \ref{alg:iq-ppo:ppos}-\ref{alg:iq-ppo:ppoe}), we use PPO \cite{schulman2017proximal} to train the policy network $\theta_\pi$ and value network $\theta_V$, both of which share a common state representation network $\theta_S$, to reduce the overall makespan. As shown in Figure \ref{fig:rl_scheduler}, we apply an MLP to compose $\theta_\pi$ and calculate the unnormalized action log probabilities (logits) of executing each pending query $q_i$ with various configurations based on the query representation $x''_i$. Then, we convert the action logits into probabilities using a softmax operation:
\begingroup
\setlength{\abovedisplayskip}{3pt}
\setlength{\belowdisplayskip}{3pt}
\begin{align*}
    \Vert_j l_i^j &= \left(\sigma \cdot \mathrm{Linear}\right)^\mathscr{p}\left(x''_i\right)\\
    \pi_\theta\left(a_i^j\vert s_t\right) &= \mathrm{softmax}\left(l_i^j\right) = \frac{\exp\left(l_i^j\right)}{\sum_{m,n} {\exp\left(l_m^n\right)}}
\end{align*}
\endgroup
where $a_i^j$ is the action to execute $q_i$ with configuration $R^j$, $l_i^j$ denotes the logit of $a_i^j$, $\pi_\theta$ is the policy, and $s_t$ is the state.

For the value network $\theta_V$, we employ another MLP to predict the state value $V(s_t)$ based on the global state representation \nolinebreak $x''_s$:
\begingroup
\setlength{\abovedisplayskip}{3pt}
\setlength{\belowdisplayskip}{3pt}
\begin{align*}
    V(s_t)=\left(\sigma \cdot \mathrm{Linear}\right)^\mathscr{q}\left(x''_s\right)
\end{align*}
\endgroup

To train the policy $\theta_\pi$ and value $\theta_V$ networks (line \ref{alg:iq-ppo:ppoe}), PPO uses a joint objective $L^{ppo}$, including the clipped surrogate objective $L^{clip}$ for $\theta_\pi$, a mean squared error loss $L^{value}$ for $\theta_V$, and an entropy bonus to ensure sufficient exploration:
\begingroup
\setlength{\abovedisplayskip}{3pt}
\setlength{\belowdisplayskip}{3pt}
\begin{align*}
    L^{clip} &= \hat{\mathbb{E}}_t \left[ \min \left( r_t(\theta) \hat{A}_t, \, \mathrm{clip} \left( r_t(\theta), 1 - \epsilon, 1 + \epsilon \right) \hat{A}_t \right) \right]\\
    L^{value} &= \hat{\mathbb{E}}_t \left[ \frac{1}{2} \left( V(s_t) - \hat{V}_t^{\mathrm{targ}} \right)^2 \right]\\
    L^{ppo} &= - L^{clip} + \beta_V L^{value} - \beta_S S[\pi]
\end{align*}
\endgroup
where $r_t(\theta) = \frac{\pi_{\theta}(a_t \mid s_t)}{\pi_{\theta_{old}}(a_t \mid s_t)}$ is the probability ratio, with $a_t$ being the action taken under $s_t$ and $\pi_{\theta_{old}}$ being the policy before updating. $\hat{A}_t$ is an estimator of the advantage function, which measures how much better $a_t$ is than others under $\pi_{\theta_{old}}\left(\cdot \mid s_t\right)$ on average. $\hat{V}_t^{\mathrm{targ}}$ is the value function target, and $\epsilon$, $\beta_V$, $\beta_S$ are hyper-parameters. The negative signs before $L^{clip}$ and $\beta_S S[\pi]$ indicate that PPO aims to maximize them during training.

During the auxiliary phase (line \ref{alg:iq-ppo:aux}), IQ-PPO enhances PPO by fully exploiting the rich signals of individual query completion, while controlling policy distortions caused by the auxiliary time prediction task. Notably, as the scheduler continuously submits new queries during scheduling, we can only predict the finish time $T_{\theta_A}(q_e\vert s_t)$ of the earliest query $q_e$ to finish among all concurrent queries at time $t$ based on the current state $s_t$. As shown in Figure \ref{fig:rl_scheduler}, we add an auxiliary network $\theta_A$, which shares the common state representation network $\theta_S$ with the policy $\theta_\pi$ and value $\theta_V$ networks, and employs an MLP to predict $T_{\theta_A}(q_e\vert s_t)$ based on the query representation $x''_e$:
\begingroup
\setlength{\abovedisplayskip}{3pt}
\setlength{\belowdisplayskip}{3pt}
\begin{align*}
    T_{\theta_A}(q_e\vert s_t) = \left(\sigma \cdot \mathrm{Linear}\right)^\mathscr{r} \left(x''_e\right)
\end{align*}
\endgroup

Corresponding to the auxiliary goal, we optimize the state representations using a joint objective $L^{joint}$, including an auxiliary loss $L^{aux}$ and a behavior cloning loss:
\begingroup
\setlength{\abovedisplayskip}{3pt}
\setlength{\belowdisplayskip}{3pt}
\begin{align*}
    L^{aux}&=\hat{\mathbb{E}}_t\left[\frac{1}{2}\left(T_{\theta_A}(q_e\vert s_t)-T_{e}^{\mathrm{targ}}\right)^2\right]\\
    L^{joint}&=L^{aux}+\beta_{clone} \cdot \hat{\mathbb{E}}_t\left[K L\left[\pi_{\theta_{old}}\left(\cdot \mid s_t\right), \pi_\theta\left(\cdot \mid s_t\right)\right]\right]
\end{align*}
\endgroup
where $T_{e}^{\mathrm{targ}}$ is the ground-truth finish time of $q_e$, and $\pi_{\theta_{old}}$ is the policy before the auxiliary phase. We use the mean squared error loss $L^{aux}$ as the first term to learn from the individual query finish time, while preserving the original policy with the second \linebreak term and controlling this trade-off via a hyper-parameter $\beta_{clone}$.

\SetAlgoSkip{SkipBeforeAndAfter}
\begin{algorithm}[t]
\caption{IQ-PPO}\label{alg:iq-ppo}
\KwIn{\mbox{Initial parameters: $\theta_S$, $\theta_\pi$, $\theta_V$, and $\theta_A$; Numbers of:} iterations $N_i$ and PPO iterations per phase $N_{ppo}$}
\KwOut{Optimized parameters: $\theta_S$, $\theta_\pi$, $\theta_V$, and $\theta_A$}
\For{i = 1, 2, \dots, $N_i$}{
    Initialize empty log $\mathbb{L}$\;
    \For(\tcp*[f]{PPO phase}){j = 1, 2, \dots, $N_{ppo}$}{\label{alg:iq-ppo:ppos}
        Schedule batch queries under current policy $\pi$ \linebreak and add the generated logs to $\mathbb{L}$\;
        Optimize $L^{ppo}$ w.r.t. $\theta_\pi$ and $\theta_V$ (with shared $\theta_S$) \linebreak on data collected in current PPO iteration\;\label{alg:iq-ppo:ppoe}
    }\vspace{-1pt}
    Extract the earliest query $q_e$ to finish and its finish \linebreak time $T_{e}^{\mathrm{targ}}$ and compute current policy $\pi_{\theta_{old}}(\cdot | s_t)$ \linebreak for all states $s_t$ in $\mathbb{L}$\;
    Optimize $L^{joint}$ w.r.t. $\theta_A$ and $\theta_\pi$ (with shared $\theta_S$) \linebreak on all data in $\mathbb{L}$ \tcp*{auxiliary phase}\label{alg:iq-ppo:aux}
}\vspace{-2pt}
\end{algorithm}

\section{Optimization Strategies}
In this section, we introduce three optimization strategies for batch query scheduling based on the RL framework above, including adaptive masking, scheduling gain-based query clustering, and an incremental simulator.

\subsection{Adaptive Masking for Action Space}
A well-designed action space significantly improves learning efficiency and assists in optimizing scheduling strategies. In each scheduling step of BQSched, we need to not only select the next query to execute from all pending queries, but also determine its running parameters to allocate proper resources to the selected query. As we discussed before, considering parameters in each step expands the scheduling space exponentially, making it difficult for RL to converge within an acceptable time cost. Specifically, since different queries have various preferences for computational resources, much time is wasted on inefficient parameter configurations during exploration, such as allocating more CPU resources to I/O-intensive queries. It is expected to improve exploration efficiency if we can identify and avoid these inefficient configurations in advance.

\begin{figure}[t]
    \centering
    \includegraphics[width=.914\linewidth]{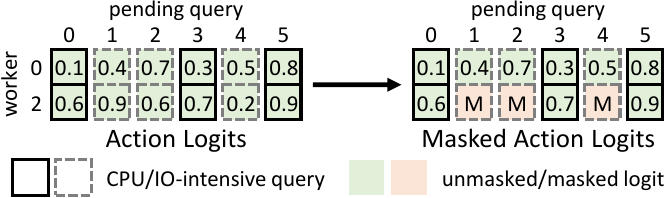}\vspace{-6pt}
    \caption{An example of adaptive masking.}
    \label{fig:adaptive_masking}\vspace{-15pt}
\end{figure}

We collect the query performance under various parameter configurations as external knowledge, and prune the available configurations for different queries adaptively, thereby shrinking the action space with limited impact on the overall efficiency. Specifically, due to the periodic nature of batch query execution, we can extract the relationships between running parameters and execution cost for each query to identify its resource preferences and guide the masking assignment. That is, we collect the absolute and relative efficiency improvements brought by allocating more resources such as workers and memory, and set thresholds to mask configurations with minor improvements. This allows us to adaptively mask inefficient configurations for each query. For these configurations, we replace the corresponding action logits with a large negative number, e.g., $M=-10^8$, indicating that the probabilities of selecting them become nearly 0 after softmax. \linebreak Taking Figure \ref{fig:adaptive_masking} as an example, we mask the action logits that allocate more workers (2) to the I/O-intensive queries (1, 2, 4).

\subsection{Scheduling Gain-Based Query Clustering}
Besides scaling with the running parameters, the space of different scheduling plans grows exponentially as the number of queries increases, leading to difficulties in RL learning within an acceptable time cost. To improve BQSched's scalability in terms of queries, we design a scheduling gain-based query clustering method and simplify the scheduling task at the cluster level.

The intuition comes from an observation that in an efficient batch query scheduling, concurrently running queries always show some performance gain, i.e., the concurrent queries usually need less time to finish in an efficient scheduling than in other schedulings. Such gain may come from more resource sharing and less resource contention among concurrent queries. If we discover queries with high scheduling gain from historical logs in advance and incorporate them into the scheduler as external knowledge, much exploration cost in RL learning will be saved.

We aim to extract the scheduling gain between queries, i.e., the performance gain brought by executing them concurrently, from batch query execution logs. These logs record multiple rounds of scheduling containing each query's start and finish time, which helps us quantify the execution efficiency of queries in scheduling. For each pair of queries, there are many cases in which the two queries run concurrently with different relative start times, and thus various execution time overlaps. Intuitively, the larger the overlap between two queries, the greater their mutual influence, and the efficiency changes better reflect their scheduling gain. Therefore, such overlaps help reveal the weight of each concurrent execution in computing the overall scheduling gain between the two queries.

\stitle{Scheduling Gain and Query Clustering.} We define the symmetric scheduling gain between two queries. Specifically, we measure the change in execution efficiency by the acceleration rate of query execution time over the average, and quantify the degree of mutual influence in each concurrent run by the proportion of overlap in query execution time. In addition, the final results are weighted based on the square root of the queries' average execution time since we focus more on complex queries. Following the above analysis, the scheduling gain between $q_i$ and $q_j$ can be computed as follows:
\begingroup
\setlength{\abovedisplayskip}{2.9pt}
\setlength{\belowdisplayskip}{2.9pt}
\begin{align*}
    gain_{ij} = gain_{ji} &= \frac{1}{m_c}\sum_{m_c}\frac{o_{ij}a_{ij}\sqrt{\overline{t}_i}+o_{ji}a_{ji}\sqrt{\overline{t}_j}}{\sqrt{\overline{t}_i}+\sqrt{\overline{t}_j}}
\end{align*}
\endgroup
where $\overline{t}_i$ is $q_i$'s average execution time in logs. For a run of batch queries during which $q_i$ and $q_j$ are executed concurrently, $a_{ij}$ is the acceleration rate of $t^j_i$ over $\overline{t}_i$, expressed as $a_{ij}=1-t^j_i/\overline{t}_i$, where $t^j_i$ is $q_i$'s execution time under the influence of $q_j$. $o_{ij}$ is $ov_{ij}$'s proportion in $t^j_i$, calculated as $o_{ij}=ov_{ij}/t^j_i$, where $ov_{ij}$ is the time overlap of $q_i$ and $q_j$'s execution. Finally, we compute the symmetric scheduling gain $gain_{ij}$ between $q_i$ and $q_j$ by averaging the weighted sum term over all $m_c$ concurrent executions of $q_i$ and $q_j$ in logs. The above design achieves a high scheduling gain between two queries when their mutual influence improves each other's execution efficiency.

Moreover, the computation of scheduling gains should have some generalization ability, as not all query pairs have been concurrently scheduled before. To this end, we use the actual gains from logs to train an MLP, which predicts the scheduling gains between queries based on their execution plan embeddings encoded by QueryFormer \cite{zhao2022queryformer} and generalizes the clustering mechanism to query pairs not covered in logs. We ensure the symmetry of the predicted values by swapping the input order of the queries and summing the results:
\begingroup
\setlength{\abovedisplayskip}{2.9pt}
\setlength{\belowdisplayskip}{2.9pt}
\begin{align*}
    \widehat{gain}_{ij}=\widehat{gain}_{ji}=\left(\sigma \cdot \mathrm{Linear}\right)^\mathscr{s}\left(e_i\Vert e_j\right)+\left(\sigma \cdot \mathrm{Linear}\right)^\mathscr{s}\left(e_j\Vert e_i\right)
\end{align*}
\endgroup

We use the scheduling gain as a similarity metric and cluster the batch queries. Since scheduling gain is either extracted from logs or fitted by an MLP, it cannot be easily represented by traditional distance metrics in a certain feature space. Therefore, we use average-linkage agglomerative clustering \cite{day1984efficient} as it operates directly on the similarity matrix without feature extraction and allows flexible control over the number of clusters. Specifically, we first initialize $n$ query clusters, each containing one query in the $n$-sized batch query set. Then, we repetitively merge two clusters according to the average scheduling gain between their respective queries in a greedy fashion, until the number of clusters decreases to the specified number $n_c$ of final clusters. Based on workload characteristics, we strike a balance between scheduling granularity and training costs by adjusting the value of $n_c$: the larger the value of $n_c$, the finer the scheduling granularity; \mbox{the smaller the value of $n_c$, the lower the training costs.}

\begin{figure}[t]
    \centering
    \includegraphics[width=.985\linewidth]{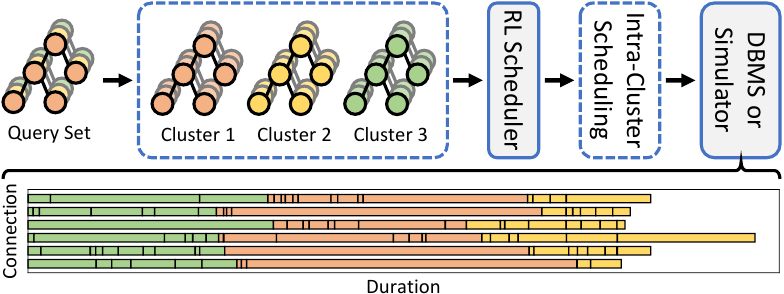}\vspace{-6.2pt}
    \caption{Scheduling on query clusters.}
    \label{fig:schedule_on_cluster}\vspace{-16.2pt}
\end{figure}

\stitle{Scheduling on Query Clusters.} Figure \ref{fig:schedule_on_cluster} shows the scheduling process on query clusters (top) and the resulting scheduling plan (bottom). The scheduling gain-based query clustering actually undertakes part of the scheduling task. That is, queries within the same cluster (horizontal bars of the same color in Figure \ref{fig:schedule_on_cluster}) are submitted to idle connections one by one to maximize concurrent execution, while the RL scheduler learns a scheduling strategy over clusters to reduce the overall makespan. For intra-cluster queries, the execution order can be determined by heuristic rules like FIFO and MCF. Due to the high scheduling gains between queries within the same cluster, these queries improve each other's execution efficiency by boosting resource sharing and reducing resource contention when executed concurrently. Thus, we reduce the adverse effects caused by the continuous submission of intra-cluster queries when we coarsen the scheduling granularity.

To adapt to the transformed scheduling granularity and learn an efficient inter-cluster scheduling strategy, we make specific extensions to the RL scheduler, including the state representation, action space, and adaptive masking. For the state representation, we obtain the representation of each cluster by sum-pooling the embeddings of its internal queries encoded by QueryFormer. For the action space, we adjust the action logits' original dimension $n$ to the number of query clusters $n_c$, representing the policy's probability of selecting each cluster. For the adaptive masking, we extend it to the query clusters with the consideration of two factors. First, all queries in the same cluster are expected to share the same parameter configuration $R^c$ to reduce the query-level parameter search space. Thus, we determine $R^c$ for each cluster by the policy network as mentioned above. Second, some queries have their own inefficient configuration mask from logs, which may conflict with the cluster-level $R^c$. For each $q_i$ in these conflicting queries, we select another configuration $R^i$, which is the closest to $R^c$ and does not conflict with the query-level mask of $q_i$.

By the above design, we reduce the size of the action space and sequence from the number of queries $n$ to the number of clusters $n_c$. Thus, the problem complexity decreases exponentially from $O\left(n! \cdot \left(\prod_{j=1}^{\left|R\right|}{k_j}\right)^n\right)$ to $O\left({n_c}! \cdot \left(\prod_{j=1}^{\left|R\right|}{k_j}\right)^{n_c}\right)$ at the cluster level. We can also adapt BQSched to various batch query set sizes by adjusting $n_c$ to balance scheduling granularity and training costs.

\subsection{Incremental Simulator for Concurrent Query Execution}
The RL scheduler still needs considerable cost in sampling directly from the DBMS, even though we have introduced strategies like IQ-PPO and adaptive masking. In this subsection, we propose a simulator to simulate feedback on query finish times from the DBMS by predicting the earliest query to finish and its finish time among all concurrent queries. This is a practical goal, as DBMSs have their own estimates of query execution costs, while the periodic nature of batch query execution ensures sufficient data from similar workloads to train the simulator effectively. Specifically, the simulator is trained over historical logs and supports the pre-training of the RL scheduler, which saves much sampling cost during training on the DBMS. In addition, the online logs generated by more recent executions can be used as incremental training data to improve the simulator's \linebreak accuracy and adaptability to the underlying data and queries.

\stitle{Concurrent Query Prediction with Shared State Representation.} To simulate feedback from the DBMS, we need to predict the finish times of individual concurrent queries based on their execution state, which is similar to the auxiliary task in IQ-PPO. However, the concurrent query prediction in the simulator requires further design, particularly to ensure that the simulator can continuously interact with the scheduler in place of the DBMS. In the auxiliary task, we only predict the finish time of the earliest query to finish to avoid excessive policy distortions. In contrast, the simulator needs to simulate the future state of concurrent queries as input for the scheduler based on their current state, which means predicting not only the finish time, but also which query finishes earliest. In this way, given all concurrent queries $S_c$ at time $t$, we can predict the earliest query to finish and its finish time $t'$ with the information at $t$. As a new query scheduled at $t'$, we can further predict the next earliest query to finish and its finish time $t''$ iteratively. Therefore, the RL scheduler can get feedback from such a simulator without actual execution, just like the underlying DBMS.

Based on the above analysis, we reuse the attention-based state representation to predict the earliest query to finish and its finish time. That is, we encode the concurrent queries as described in Section \ref{subsec:attention_based_state_representation} and add two final layers for the prediction tasks. Specifically, we use an MLP and a softmax operation to predict each concurrent query $q_i$'s probability $p_i$ to be the earliest query to finish based on the batch query representation $x''_i$. Besides, we use another MLP to predict the earliest finish time $T_e$ based on the representation of the predicted earliest query to finish:
\begingroup
\setlength{\abovedisplayskip}{4.3pt}
\setlength{\belowdisplayskip}{4.3pt}
\begin{align*}
    p_i &= \mathrm{softmax}\left(\left(\sigma \cdot \mathrm{Linear}\right)^\mathscr{t}\left(x''_i\right)\right)\\
    T_e &= \left(\sigma \cdot \mathrm{Linear}\right)^\mathscr{u}\left(x''_{\arg\max_i\left(p_i\right)}\right)
\end{align*}
\endgroup

Then, we introduce multitask learning \cite{caruana1997multitask} to optimize both objectives simultaneously. We define the overall loss $L^{oa}$ as the weighted sum of cross-entropy loss $L^{clf}$ for the classification task and mean squared error loss $L^{reg}$ for the regression task:
\begingroup
\setlength{\abovedisplayskip}{4.3pt}
\setlength{\belowdisplayskip}{4.3pt}
\begin{gather*}
    L^{clf} = \mathrm{CE}\left(id^{\mathrm{targ}}, p\right),\ L^{reg} = \mathrm{MSE}\left(T_e^{\mathrm{targ}}, T_e\right)\\
    L^{oa} = L^{clf} + \gamma \cdot L^{reg}
\end{gather*}
\endgroup
where $id^{\mathrm{targ}}$ is the ground-truth index of the earliest query to finish, and $\mathrm{CE}$ is the cross entropy. $T_e^{\mathrm{targ}}$ is the ground-truth finish time of the earliest query to finish, and $\mathrm{MSE}$ is the mean squared error. $\gamma$ is the scaling coefficient of the regression loss.

\begin{table*}[htbp]
\caption{Efficiency ($\overline{t}^{ov}$/s) and stability ($\sigma^{ov}$/s) on three benchmarks with three databases.}\vspace{-6pt}
\label{tab:cross_database_benchmark}
\centering
\begin{tabular}{@{\extracolsep{2pt}}cccccccccc} \toprule
\multirow{2}{*}{Strategy} & \multicolumn{3}{c}{DBMS-X}                                            & \multicolumn{3}{c}{DBMS-Y}                                 & \multicolumn{3}{c}{DBMS-Z}                                           \\ \cline{2-4} \cline{5-7} \cline{8-10}
                          & TPC-DS                  & TPC-H                  & JOB                    & TPC-DS                  & TPC-H                  & JOB                    & TPC-DS                  & TPC-H                  & JOB                     \\ \midrule
Random                    & 20.71$\pm$1.68          & 6.17$\pm$0.94          & 9.75$\pm$0.69          & 20.11$\pm$2.17          & 4.97$\pm$0.41          & 7.24$\pm$0.32          & 8.68$\pm$0.84          & 1.07$\pm$0.12          & 8.49$\pm$0.61          \\
FIFO                      & 20.05$\pm$1.36          & 6.26$\pm$0.06          & 10.57$\pm$0.21         & 16.90$\pm$2.60          & 5.91$\pm$0.29          & 7.14$\pm$0.18          & 9.04$\pm$0.13          & 1.07$\pm$0.04          & 8.99$\pm$0.11          \\
MCF                       & 19.01$\pm$1.54          & 5.05$\pm$0.61          & 8.78$\pm$0.22          & 15.01$\pm$3.68          & 4.93$\pm$0.22          & 7.12$\pm$0.09          & 7.37$\pm$0.10          & 0.90$\pm$0.07          & 8.19$\pm$0.07          \\
LSched                    & 16.91$\pm$0.57          & 4.64$\pm$0.04          & 8.50$\pm$0.15          & 12.03$\pm$2.27          & 3.74$\pm$0.13          & 6.82$\pm$\textbf{0.05} & 7.26$\pm$0.07          & 0.84$\pm$\textbf{0.02} & 8.07$\pm$0.07          \\
BQSched                   & \textbf{14.39$\pm$0.09} & \textbf{3.65$\pm$0.03} & \textbf{7.96$\pm$0.03} & \textbf{10.45$\pm$0.37} & \textbf{3.59$\pm$0.12} & \textbf{6.80$\pm$0.05} & \textbf{7.01$\pm$0.06} & \textbf{0.76$\pm$0.02} & \textbf{7.83$\pm$0.04} \\ \bottomrule
\end{tabular}\vspace{-14pt}
\end{table*}

\stitle{RL Training with Simulator.} To reduce the sampling cost, we pre-train the RL scheduler by having it interact with the simulator instead of the DBMS. Given a batch query set, the scheduler first uses its current policy to determine the initial queries \linebreak and submits them to the simulator. The simulator then predicts the earliest query to finish and its finish time based on the current state. After the simulator sends back the query finish signal, the scheduler selects the next query and submits it to the simulator again. Such interactions will continue until all the batch queries finish, which constitute a simulated sequence of scheduling.

Due to the accumulated errors from the simulator and the potential changes in the underlying data, we further fine-tune the RL scheduler on the underlying DBMS after pre-training. Specifically, we first save multiple intermediate scheduler models in the pre-training phase. Then, we verify all the saved models on the actual database, and choose the most efficient one for fine-tuning. Finally, we continue to train the scheduler and fine-tune its scheduling strategy on the DBMS. Such a pretrain-finetune paradigm significantly reduces the time cost and resource consumption compared with training on the underlying DBMS from scratch. Moreover, if performance degradation is observed during the deployment of our fine-tuned strategy, the scheduler can be further fine-tuned with new interaction data collected online to adapt to edge cases and improve robustness during inference. The batch query execution logs continuously generated by online fine-tuning and inference also provide incremental training data for the prediction model, thereby improving our simulator's accuracy and ability to adapt to potential changes in the underlying data and queries over time.

\section{Experimental Evaluation}
In this section, we describe the experimental setup, compare BQSched with other methods across various scenarios, show the effectiveness of different components in BQSched, and visualize the learned scheduling plan in a case study.

\subsection{Experimental Setup}
\stitle{Benchmarks.} We conduct experiments on three benchmarks:

\textit{TPC-DS} \cite{nambiar2006making} contains 99 query templates. We conduct most experiments on TPC-DS as it best matches our scenario. We use the official generator to generate data with scale factors of 1, 2, 5, 10, 50, 100, and 200. We generate 1x, 2x, 5x, and 10x queries and take all 99, 198, 495, and 990 queries as the input batch query sets. Moreover, we make the following extensions to TPC-DS: 1) We optimize the templates of queries 1, 6, 30, and 81 without affecting the results, as their original execution is extremely inefficient. 2) We generate 0.8x, 0.9x, 1.1x, and 1.2x data and queries by discarding or duplicating the corresponding portions of \linebreak the original data and queries to test the adaptability of BQSched.

\textit{TPC-H} \cite{poess2000new} contains 22 query templates. We use the official generator to generate data with scale factors of 1, 50, 100, and 200, and form the batch query set with all 22 queries.

\textit{JOB} \cite{leis2015good}, based on the IMDb dataset, contains 113 queries generated from 33 templates. We take the first query generated from each template, i.e., 1a to 33a, to build the batch query set, as scheduling similar queries is rare in real-world applications.

\stitle{Competitors.} Existing pipeline building tools, like DBT \cite{cyr2023introduction}, follow simple heuristics in query scheduling, such as Random and FIFO. We can also use MCF to mitigate the long-tail query problem when the query execution time can be extracted from logs. Moreover, we adapt the existing RL-based scheduler \cite{sabek2022lsched} to our context by migrating the state representation in BQSched, as we cannot find another non-intrusive RL scheduler for batch queries. The competitors in this paper are as follows:
\begin{itemize}[wide, labelwidth=!, labelindent=0pt]
    \item \textit{Random} schedules queries in a random order.
    \item \textit{FIFO} (First In, First Out) schedules queries in their submission order.
    \item \textit{MCF} (Maximum Cost First) schedules queries according to their average execution time in logs from long to short.
    \item \textit{LSched} \cite{sabek2022lsched} is an operator-level RL-based scheduler for analytical database systems. To support query-level scheduling, we make necessary modifications to LSched. Firstly, we exclude the features that cannot be collected outside the underlying DBMS, and include the state representation of BQSched as a supplement. Secondly, we change the policy network in LSched to predict the probability of selecting each query and its corresponding running parameters, as in BQSched.
\end{itemize}

\stitle{Environment Settings.} We conduct experiments on three underlying database systems: DBMS-X, Y, and Z, as there are various data sources in data pipelines. For DBMS-X, we run it on an Ubuntu 18.04 server with two Intel(R) Xeon(R) Gold 5218 CPUs and 256GB memory. We keep the default settings except for disabling the nested loop join operator, which may result in poor plans for some queries in TPC-DS. DBMS-Y is on an Ubuntu 20.04 server with two Intel(R) Xeon(R) Gold 5318Y CPUs and 256GB memory. DBMS-Z contains three computing nodes, each with 16 vCPUs and 64GB memory. We implement BQSched with Python 3.8, Pytorch 2.0 for learning, and Stable Baselines 1.8 as the RL framework. We train BQSched on an Ubuntu 20.04 server with two Intel(R) Xeon(R) Gold 5318Y CPUs, 256GB memory, and one NVIDIA RTX A5000 GPU.

\stitle{Implementation Details.} We configure the hyper-parameters of all baselines according to their original papers. For BQSched, we keep the default hyper-parameters of PPO \cite{schulman2017proximal}, train the policy and value networks every 25 rounds of scheduling, and evaluate the model every 50 rounds of scheduling. We set $N_i$ large enough to ensure RL convergence and set $N_{ppo}$ to 10.

\begin{table*}[htbp]
\caption{Efficiency ($\overline{t}^{ov}$/s) and stability ($\sigma^{ov}$/s) when data and query set change on TPC-DS with DBMS-X.}\vspace{-6.6pt}
\label{tab:adapt_result}
\centering
\begin{tabular}{@{\extracolsep{2pt}}ccccccccc} \toprule
\multirow{2}{*}{Strategy} & \multicolumn{4}{c}{Data Scale}                                                                        & \multicolumn{4}{c}{Query Scale}                                                                       \\ \cline{2-5} \cline{6-9}
                          & 0.8x                    & 0.9x                    & 1.1x                    & 1.2x                    & 0.8x                    & 0.9x                    & 1.1x                    & 1.2x                    \\ \midrule
Random                    & 16.26$\pm$0.93          & 19.48$\pm$1.91          & 23.79$\pm$0.98          & 26.59$\pm$1.61          & 20.66$\pm$1.15          & 20.65$\pm$1.37          & 22.20$\pm$1.29          & 23.92$\pm$2.43          \\
FIFO                      & 15.30$\pm$1.17          & 17.86$\pm$0.78          & 25.82$\pm$0.23          & 28.30$\pm$0.15          & 20.23$\pm$1.67          & 19.90$\pm$2.13          & 20.95$\pm$1.92          & 23.95$\pm$2.17          \\
MCF                       & 15.41$\pm$1.31          & 17.59$\pm$1.56          & 22.28$\pm$0.25          & 23.95$\pm$0.17          & 20.59$\pm$1.29          & 19.36$\pm$1.90          & 22.39$\pm$3.33          & 21.51$\pm$1.17          \\
LSched                    & 13.48$\pm$1.02          & 15.36$\pm$1.04          & 24.84$\pm$1.35          & 26.56$\pm$1.90          & 16.95$\pm$0.24          & 17.39$\pm$1.27          & 18.27$\pm$1.45          & 19.59$\pm$1.38          \\
BQSched                   & \textbf{12.88$\pm$0.16} & \textbf{13.95$\pm$0.15} & \textbf{21.81$\pm$0.20} & \textbf{23.69$\pm$0.14} & \textbf{14.34$\pm$0.09} & \textbf{14.67$\pm$0.18} & \textbf{14.88$\pm$0.20} & \textbf{15.59$\pm$0.07} \\ \bottomrule
\end{tabular}\vspace{-15pt}
\end{table*}

\subsection{Batch Query Scheduling}

\stitle{Efficiency.} Table \ref{tab:cross_database_benchmark} shows the average makespans of different strategies across three benchmarks on different DBMSs. Among the heuristics, Random and FIFO have similar performance with low efficiency, and MCF has a slight improvement. For the learned strategies, LSched outperforms the heuristics but still lags behind BQSched, partly due to the limited capabilities of its RL algorithm. BQSched achieves the best performance on all benchmarks and DBMSs. On TPC-DS and TPC-H, BQSched brings an average performance improvement of 30\% to 37\% over Random and FIFO, and outperforms LSched by 11\% on average. Even on JOB with relatively limited optimization space, BQSched learns the most efficient scheduling strategy stably and improves FIFO's efficiency by about 14\%.

Table \ref{tab:cross_database_benchmark} also shows the performance of different strategies across three DBMSs. BQSched decouples the query execution and scheduling components, treats the DBMS as a black box, and thus supports various DBMSs easily. The results show that BQSched achieves the best performance on DBMS-X, Y, and Z compared with other methods. Among them, DBMS-X has the largest scheduling potential, on which BQSched achieves an average efficiency improvement of 32\%, 20\%, and 14\% over FIFO, MCF, and LSched, respectively. In addition, BQSched outperforms FIFO by about 21\% on DBMS-Z under 1x data, which is not as significant as those on the other DBMSs. We speculate that DBMS-Z has its own scheduling mechanism to improve resource utilization in concurrent query execution, leading to \linebreak limited scheduling potential for BQSched and other methods.

\stitle{Stability.} Table \ref{tab:cross_database_benchmark} also shows the stability of different methods. All scheduling strategies exhibit some deviations due to uncertainty in concurrent execution. The results show that the heuristic strategies lead to significantly large $\sigma^{ov}$, which means greater fluctuations, especially on TPC-DS. This is partly because the heuristics lack modeling of the complex query patterns. In contrast, the RL-based methods can implicitly capture these patterns from feedback, thereby improving policy stability. As the results show, BQSched improves the stability of FIFO, MCF, and LSched by 68\%, 68\%, and 38\% on average, indicating that the scheduler improves its resistance to potential state disturbances \linebreak while learning to allocate concurrent resources during training.

\stitle{Scalability.} We conduct scalability tests from both the data and query perspectives.

\begin{figure}[t]
    \centering
    \includegraphics[width=\linewidth]{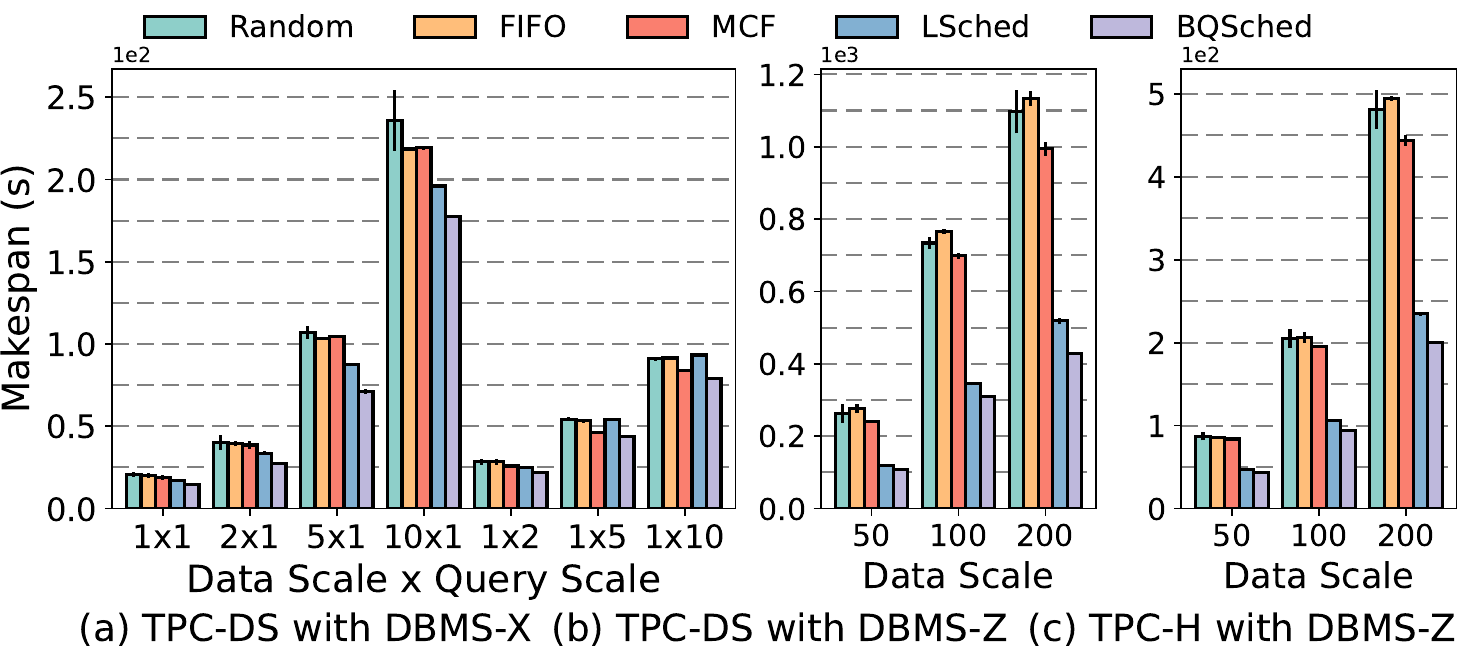}\vspace{-6.6pt}
    \caption{Scalability on TPC-DS and TPC-H with DBMS-X and Z.}
    \label{fig:scale_result}\vspace{-16.2pt}
\end{figure}

Figure \ref{fig:scale_result}(a) shows the results on TPC-DS with DBMS-X under 1x, 2x, 5x, and 10x data. As the data scale increases, the efficiency of MCF gradually drops and becomes even lower than FIFO under 5x and 10x data. We speculate that as MCF schedules the queries with high costs early, the resource contention among these queries becomes a major issue when data increases on the DBMS with fixed resources. In contrast, the RL-based methods show stable improvement, as they can be aware of resource contention and perform adaptive optimization for various underlying data. For example, BQSched outperforms FIFO by \linebreak 30\%, 31\%, and 19\% under 2x, 5x, and 10x data, respectively.

Figures \ref{fig:scale_result}(b) and \ref{fig:scale_result}(c) increase the data scale factor to 50, 100, and 200, and show the results on TPC-DS and TPC-H, respectively. As the performance of the centralized DBMS-X significantly drops when data increases, we conduct experiments on the distributed DBMS-Z to better align with real-world applications under large data scales. The results show that the distributed environment somewhat mitigates the resource contention issue of MCF, which outperforms FIFO by 9\% on average. However, even though DBMS-Z achieves the best performance among the three DBMSs, its internal scheduling mechanism still cannot efficiently schedule queries in TPC-DS and TPC-H under large data scales, thereby providing significant scheduling potential for the RL-based schedulers. Specifically, BQSched improves FIFO's efficiency by 55\%, 57\%, and 61\% under 50x, 100x, and 200x data, respectively.

Figure \ref{fig:scale_result}(a) also shows the results on TPC-DS with DBMS-X under 1x, 2x, 5x, and 10x queries. The efficiencies of different heuristic strategies are relatively unaffected as the query scale increases. However, the increasing scale of the query set imposes a huge impact on the RL-based methods as the scheduling space grows exponentially, leading to difficulties in RL learning. LSched even fails to learn a strategy better than the heuristics under 5x and 10x queries. In contrast, BQSched still learns a more efficient strategy than all baselines owing to the scheduling gain-based query clustering. Specifically, BQSched outperforms FIFO by 23\%, 18\%, and 13\% under 2x, 5x, and 10x queries.

\stitle{Adaptability.} Table \ref{tab:adapt_result} shows the adaptability of different strategies when the underlying data and query set change. Such an extension does not affect the heuristics but may pose challenges to the RL-based methods. That is, we train an RL scheduler on one dataset and directly apply the learned strategy to another dataset. The discrepancy between the source and target datasets may affect the scheduling performance. The results show that BQSched has sufficient adaptability to mitigate the performance degradation due to minor changes in the data and query set. When the data scale changes, BQSched achieves average efficiency improvements of 17\% and 9\% over FIFO and LSched. When the query scale changes, the improvements are 30\% and 18\%, respectively. Such adaptability may come from the single query representation, which is built upon QueryFormer \linebreak and considers both the underlying data and query features.

\stitle{Robustness.} As RL is known to cause instability during inference, it is crucial to understand the potential worst-case performance of BQSched. Unlike other RL applications such as query optimization, RL-based non-intrusive query scheduling operates above the DBMS's internal resource management and can thus naturally avoid catastrophic outcomes. In our experiments, the performance improvement of the best scheduling strategy over the worst is no more than 63\% across all scenarios. In other words, even if the RL scheduler falls back to a random strategy in the worst case, it will not produce extremely poor plans with unacceptable performance degradation.

\begin{figure}[t]
    \centering
    \includegraphics[width=\linewidth]{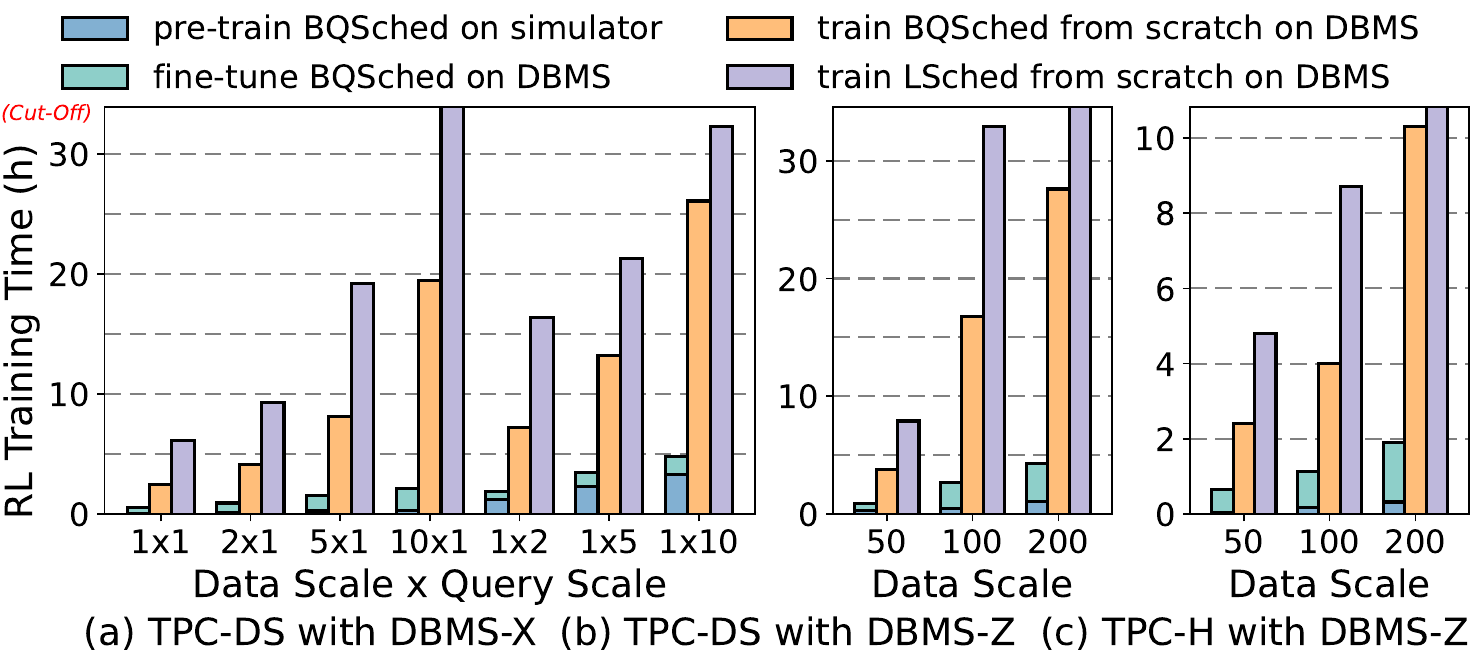}\vspace{-8.2pt}
    \caption{Training Cost on TPC-DS and TPC-H with DBMS-X and Z.}
    \label{fig:ablation_simulator}\vspace{-10.19pt}
\end{figure}

\stitle{Training Cost.} Figure \ref{fig:ablation_simulator} shows the training times of BQSched and LSched under various data and query scales on TPC-DS and TPC-H with DBMS-X and Z. The preparation steps before scheduler training, including database statistics collection, external knowledge generation, scheduling gain calculation, and simulator training, have relatively fixed time costs and take no more than 15 minutes in total, so are not included. Although BQSched has a more complex structure, it optimizes sample generation and utilization via multiple strategies, thus reducing RL training time. The results show that training BQSched takes on average only 10\% of LSched's training time across various data scales, and only 47\% even without the simulator. Notably, as the query scale increases, the search space becomes so large that LSched falls into a local optimum after a period of training and struggles to learn a strategy better than the heuristics, as shown in Figure \ref{fig:scale_result}. Therefore, LSched's training time becomes close to that of BQSched without the simulator.

\begin{figure}[t]
    \centering
    \includegraphics[width=\linewidth]{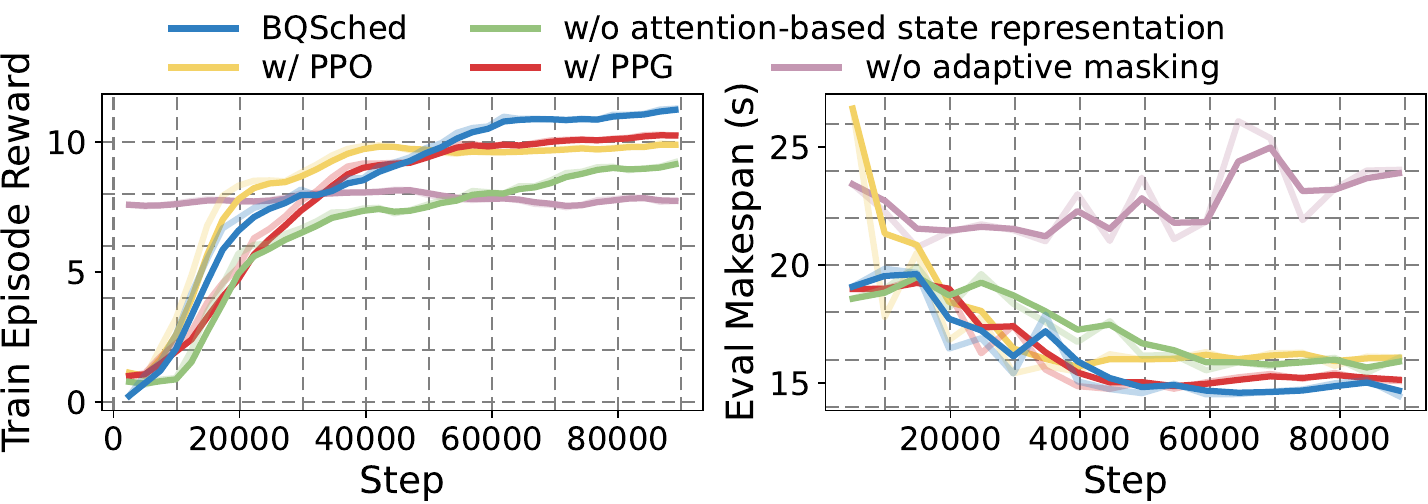}\vspace{-8.2pt}
    \caption{Ablation study of the RL scheduler and adaptive masking.}
    \label{fig:ablation_study}\vspace{-16.8pt}
\end{figure}

\subsection{Ablation Study}

\stitle{RL Scheduler.} We separately remove the attention-based state representation and IQ-PPO, and train on TPC-DS with DBMS-X from scratch to better understand their contributions.

The green line in Figure \ref{fig:ablation_study} shows BQSched's performance  after removing the attention-based state representation and concatenating the running state features of each query as an alternative. While this variant still learns a relatively efficient strategy, such a modification leads to a reduction in efficiency by about 7\%. Moreover, the strategy learned by this variant has insufficient adaptability to other query sets, as the model is too sensitive \linebreak to the query positions if we just concatenate the features directly.

The yellow and red lines in Figure \ref{fig:ablation_study} show the results when BQSched uses PPO \cite{schulman2017proximal} and PPG \cite{cobbe2021phasic} to optimize the policy and value networks after removing IQ-PPO. Due to limited sample utilization, PPO produces a strategy with relatively poorer performance within a similar training time, even though it learns slightly faster in the early stage. PPG employs an auxiliary task like BQSched, but its reuse of the estimated state values leads to a performance that, although better than PPO, still lags behind BQSched. In contrast, IQ-PPO enhances sample utilization by fully exploiting the rich signals of individual query completion, thereby improving the efficiency and stability of policy learning.

\stitle{Adaptive Masking.} The purple line in Figure \ref{fig:ablation_study} shows BQSched's performance without adaptive masking. We employ the same settings as in the RL scheduler's ablation study. Due to the expansion of action space, the scheduler fails to fully utilize the running parameters to improve resource allocation among concurrent queries within a limited time cost. From the results, the learned strategy shows a performance degradation of about 44\%, which is even worse than the heuristic strategies.

\begin{figure}[t]
    \centering
    \includegraphics[width=\linewidth]{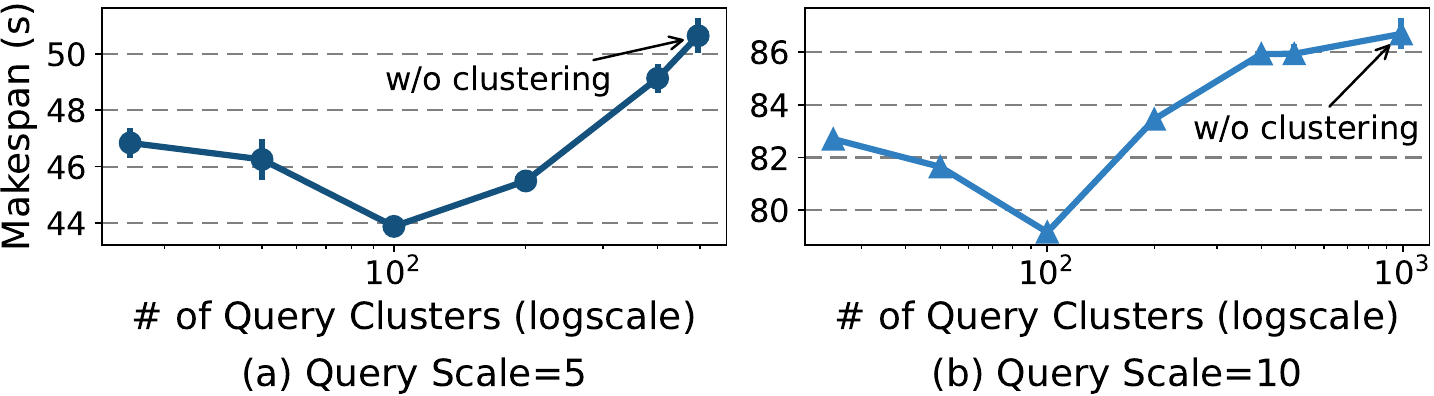}\vspace{-7pt}
    \caption{Parameter sensitivity of query clustering.}
    \label{fig:param_sensitivity}\vspace{-9.3pt}
\end{figure}

\begin{table}[t]
\caption{Ablation study and parameter sensitivity of the simulator's prediction model.}\vspace{-6pt}
\label{tab:ablation_parameter_simultor}
\centering
\begin{tabular}{cccccc} \toprule
Metric & w/o Att & w/o MTL & $\gamma=0.01$ & $\gamma=0.1$         & $\gamma=1$  \\ \midrule
Acc    & 56.6\%  & 58.6\%  & 64.4\%   & \textbf{68.7\%} & 68.5\% \\
MSE    & 0.180   & 0.102   & 0.115    & \textbf{0.073}  & 0.173  \\ \bottomrule
\end{tabular}\vspace{-14pt}
\end{table}

\stitle{Query Clustering.} Figure \ref{fig:param_sensitivity} shows the effectiveness of query clustering and the impact of the cluster number $n_c$ on TPC-DS with DBMS-X. We use scenarios with various $n_c$ values under 5x and 10x queries. As the query scale increases, the scheduling space grows exponentially, which prevents the RL scheduler from learning efficient strategies. In contrast, scheduling gain-based query clustering improves the efficiency of the learned strategies by 13\% and 9\% under 5x and 10x queries, respectively. In addition, we should select an appropriate value for $n_c$ based on workload characteristics to strike a balance between scheduling space and training costs. The results show that setting $n_c=100$ achieves the best scheduling performance in our scenario.

\stitle{Incremental Simulator.} Figure \ref{fig:ablation_simulator} also shows the impact of the simulator on BQSched's training time. The time cost of training on the DBMSs from scratch is unacceptable, especially under large amounts of data and queries. In contrast, by introducing the two-stage training with the simulator, we achieve similar performance across all scales, while spending on average only about 6\% and 15\% of the original training time on pre-training and fine-tuning, respectively. Moreover, as the data and query scales increase, the training efficiency improvement brought by the simulator becomes more evident, revealing BQSched's potential to handle larger amounts of data and queries.

\begin{figure*}
    \centering
    \includegraphics[width=\linewidth]{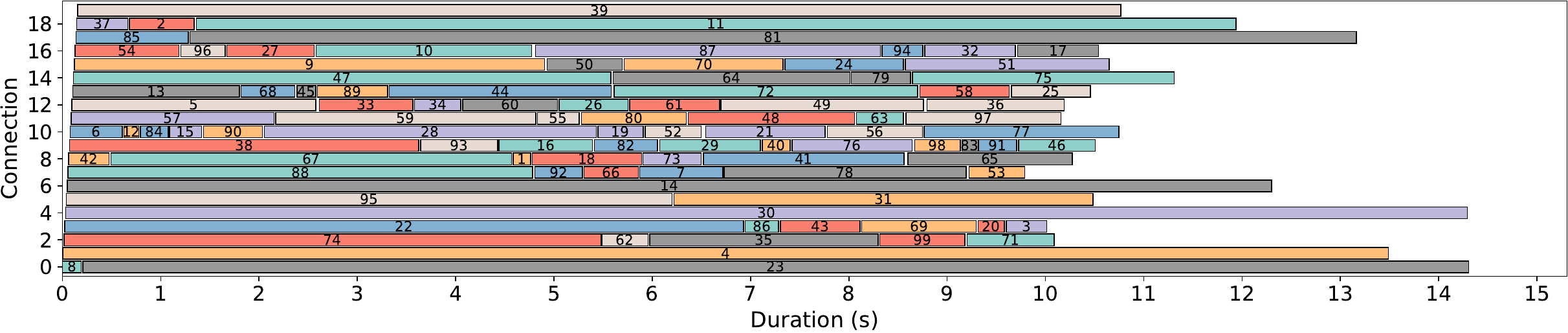}\vspace{-10.4pt}
    \caption{A case of a batch query scheduling plan learned by BQSched on TPC-DS with DBMS-X.}
    \label{fig:exec_stat}\vspace{-19.23pt}
\end{figure*}

Table \ref{tab:ablation_parameter_simultor} verifies the effectiveness of the attention-based state representation (Att) and multitask learning (MTL) in the simulator's prediction model via classification accuracy (Acc) and mean squared error (MSE) in regression. Both components enhance the prediction performance. Specifically, Att effectively captures complex query patterns, while MTL introduces additional domain information for different tasks. In addition, we explore the impact of the scaling coefficient $\gamma$ in MTL. The results show that it is necessary to fine-tune $\gamma$ based on specific scenarios to achieve optimal prediction performance, e.g., $\gamma= \linebreak 0.1$ under 1x data and 1x queries on TPC-DS with DBMS-X.

\subsection{Case Study}
Figure \ref{fig:exec_stat} visualizes a scheduling plan by BQSched under 1x data and 1x queries on TPC-DS with DBMS-X, where the numbers on the horizontal bars are query IDs. BQSched aims to maximize resource sharing, minimize resource contention, and avoid the long-tail query problem. The strategy learned by BQSched is different from the heuristics like MCF. For example, BQSched not only learns to submit complex queries such as 4, 14, and 39 as early as possible, but also combines simple queries like 6, 8, and 37 to balance concurrent resource consumption. In fact, it is almost impossible to design such a strategy via explicit rules \linebreak due to the difficulty of expressing the complex query patterns.

\section{Related Work}

\stitle{Single and Concurrent Query Modeling.} It is crucial to model SQL queries carefully before learning an efficient scheduling strategy. Recent methods for single query modeling have been proposed to support tasks like cost estimation by considering both query features and data distribution \cite{sun2019end, hilprecht2020deepdb, yang2020neurocard, wu2021unified}. Among them, QueryFormer \cite{zhao2022queryformer} achieves state-of-the-art performance with a tree Transformer \cite{vaswani2017attention}. In addition, recent works \cite{wu2013towards, zhou2020query} consider the influences among physical operators, like data sharing and resource contention, to improve performance prediction for concurrent queries. As BQSched is designed to be non-intrusive, we choose QueryFormer to model the physical plan of each query as a whole, and build the overall state \linebreak representation for concurrent queries via multi-head attention.

\stitle{RL Backbones.} An RL agent learns the value function, policy, or both through multiple rounds of interaction with the environment, known as training episodes \cite{arulkumaran2017deep}. The value function estimates the cumulative reward starting from a given state or state-action pair, while the policy directly maps states to actions. RL learning can be either on-policy or off-policy: on-policy methods \cite{singh2000convergence, mnih2016asynchronous} update the policy only based on actions taken by the current policy, while off-policy methods \cite{watkins1992q, mnih2015human} have no such restriction. PPO \cite{schulman2017proximal}, which directly optimizes the policy and value networks in an on-policy fashion, serves as the default RL algorithm in OpenAI and achieves outstanding performance in different applications. Although PPO strikes a balance between exploration and exploitation, its on-policy nature limits sample utilization, especially in scenarios with sparse feedback and high sampling cost, such as batch query scheduling. PPG \cite{cobbe2021phasic}, a PPO variant, improves sample utilization via auxiliary training with off-policy data. Some works \cite{jaderberg2017reinforcement, lee2019making} also design auxiliary tasks for various scenarios to assist in learning shared state representations. Other works \cite{todorov2012mujoco, mnih2013playing, mao2019learning} reduce sampling cost via simulators built from external knowledge.

\stitle{RL-Based Scheduling.} Recently, RL-based methods have become a hot topic in various scheduling tasks, such as job scheduling in data processing clusters \cite{mao2019learning} and operator-level query scheduling in database systems \cite{sabek2022lsched}. As mentioned before, we cannot directly apply these methods due to the complex query patterns, large scheduling space, high sampling cost, and poor sample utilization in batch query scheduling. Other works \cite{peng2021dl2, chen2022rifling, xing2023dual} use RL to schedule distributed DL jobs and reduce training time. However, they assume that DL jobs have relatively fixed resource requirements and almost no resource sharing, which differs from our scenario. Broadly, the formulated scheduling problem can be seen as a special case of combinatorial optimization, where RL has also played a vital role recently. For example, Kool et al. \cite{kool2018attention} propose an attention-based model and optimize it with RL to solve routing problems, which \linebreak inspires us with the state representation design in BQSched.

\vspace{-.68pt}\section{Conclusion}\vspace{-.68pt}
In this paper, we propose BQSched, a non-intrusive scheduler for batch concurrent queries via reinforcement learning. We first introduce an RL-based batch query scheduler, including the attention-based state representation to capture complex query patterns and the IQ-PPO algorithm to improve sample utilization. We then propose three optimization strategies, including the adaptive masking to prune the action space, the scheduling gain-based query clustering to deal with large query sets, and the incremental simulator to reduce sampling cost. Experimental results across various benchmarks, databases, data scales, and query scales show that BQSched achieves remarkable efficiency, stability, scalability, and adaptability in batch query scheduling, and outperforms the heuristic and RL-based strategies. Generally speaking, the advance of BQSched provides insights into speeding up pipeline execution safely without additional hardware or service expenditure.

\vspace{-.68pt}\section*{Acknowledgments}\vspace{-.68pt}
This work is supported by NSFC (No. 62272008) and Huawei Cloud-PKU joint program (No. TC20221205051).

\bibliographystyle{IEEEtran}
\bibliography{reference}

\end{document}